\documentclass[%
 aip,
 amsmath,amssymb,
 reprint,%
]{revtex4-1}

\usepackage{graphicx}% Include figure files
\usepackage{dcolumn}% Align table columns on decimal point
\usepackage{bm}% bold math
\usepackage[utf8]{inputenc}
\usepackage[T1]{fontenc}
\usepackage{mathptmx}
\usepackage{etoolbox}

\usepackage[inkscapelatex=false]{svg}
\usepackage{alphabeta}
\usepackage{makecell}

\makeatletter
\def\@email#1#2{%
 \endgroup
 \patchcmd{\titleblock@produce}
  {\frontmatter@RRAPformat}
  {\frontmatter@RRAPformat{\produce@RRAP{*#1\href{mailto:#2}{#2}}}\frontmatter@RRAPformat}
  {}{}
}%
\makeatother
\begin{document}

\preprint{AIP/123-QED}

\title[80 Channel Photon Pair Source from a Thin-Film Lithium Niobate Racetrack Microresonator]{80 Channel Photon Pair Source from a Thin-Film Lithium Niobate Racetrack Microresonator}
% Force line breaks with \\
\author{Mihir Chaudhari}
\author{Ian Christen}%
\author{Xinyi Ren}%
\author{Chun-Ho Lee}%
\author{Tushar Sanjay Karnik}%
\author{Reshma Kopparapu}%
\author{Clayton Cheung}%
\author{Kai-Chi Chang}%
\author{Mengjie Yu}
 \homepage{Electronic mail: mengjie.yu@berkeley.edu}
\affiliation{ 
Department of Electrical Engineering and Computer Sciences, University of California, Berkeley, Berkeley, CA 94720, USA
}%

\date{13 June 2026}% It is always \today, today,

\begin{abstract}
Nanophotonic platforms have multiple properties and features desirable for producing correlated photon pairs. In these platforms, spontaneous parametric down-conversion (SPDC) and spontaneous four-wave-mixing (SFWM) have been used to achieve photon pair production. By placing the nonlinear region in a resonant cavity, the source obtains an enhanced nonlinear interaction and naturally filters out photons by frequency according to the cavity linewidth. Among various platforms, X-cut thin-film lithium niobate (TFLN) stands out for its strong second-order optical nonlinearity ($\chi^{(2)}$) and access to a strong platform for modulating light. Although previously demonstrated, present sources are limited in their photon pair channel availability, limiting the potential of quantum applications in quantum information processing, communication, and sensing that can scale in robustness by having access to multiple photon pair channels. To this end, we demonstrate a photon pair source on X-cut TFLN with 80 measured photon pair channels spanning the C and L optical bands with a frequency spacing of 49.5 GHz between photons. We characterize the second-order correlation function and pair generation rate (PGR) of 80 channels, the highest number of channels demonstrated to date, and achieve a pair generation of 125.7 kHz / μW after accounting for the cavity's escape efficiency. We also demonstrate heralded single photon operation by calculating a heralded auto-correlation dip of 0.0544 $\pm$ 0.0054 with 2.2 mW of on-chip pump power. These findings demonstrate the promise of X-cut TFLN as a strong platform for quantum optical applications.
\end{abstract}

\maketitle

\section{Introduction}

\begin{figure*}
\includesvg[width=1\textwidth]{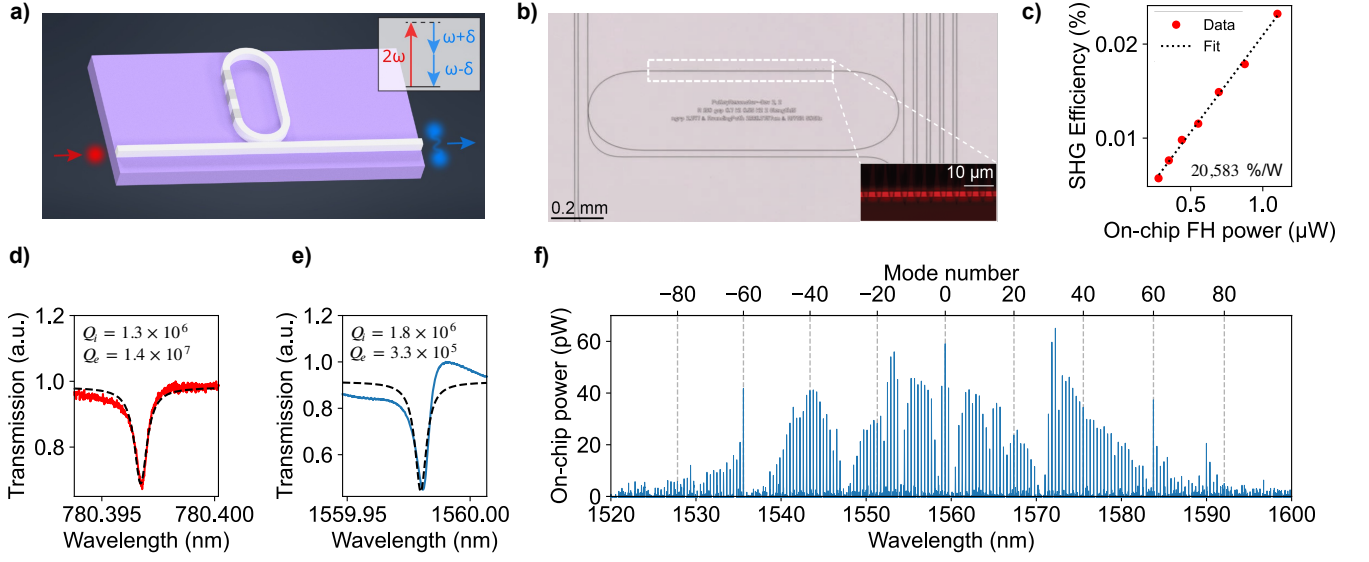}
\caption{\label{fig:wide}Device design and performance. (a) Multi-channel photon pair device schematic. (b) Microscope image of the fabricated racetrack microresonator on X-cut TFLN. The inset is a SHG confocal image of the periodically-poled region. (c) Normalized SHG efficiency fitted to be 20,583 \%/W. (d) Visible transmission. (e) Telecom transmission. (f) Downconverted spectrum using 26.2 mW of on-chip visible power near 780 nm.}
\end{figure*}

Photon pair sources are the foundation for applications such as quantum key distribution (QKD), quantum secure direct communication (QSDC), quantum teleportation, discrete-variable (DV) systems, continuous-variable (CV) systems, and quantum imaging \cite{quant_comm, quant_teleportation_2023, quant_compute, quant_imaging}. In QKD and QSDC, entangled photon pairs serve as a resource for secure communication protocols \cite{long_distance_qkd_2022, qsdc_2021}. Quantum teleportation requires entangled pairs to provide non-classical correlations for transferring unknown quantum states when combined with classical communication channels \cite{quant_teleportation_1997}. DV systems can use heralded single photons for quantum computations such as boson sampling \cite{time_multiplex_herald, boson_sampling}. CV systems rely on the generation of squeezed light, which can be generated using the same processes that create photon pairs \cite{cv_quantum_field_theory, cv_neural_networks, cv_architecture}. Lastly, quantum imaging can use photon pairs for different schemes, including ghost imaging, heralded imaging, biphoton imaging, and undetected photon imaging \cite{ghost_imaging, heralded_imaging, biphoton_imaging, undetected_photon_imaging}.

While entangled photon pairs enable numerous quantum applications, access to multiple photon pair channels can make these applications more capable and robust. For example, QKD supports more users with more channels \cite{multi_channel_qkd, multiplex_communication, multiplex_communication_2025}. Communication channel capacity can be increased with multiplexing \cite{multiplex_communication_rate_2012, multiplex_channel_capacity_improvement_2020, multiplex_channel_capacity_improvement_2021}. Multiplexing can also improve the reliability of communication by improving fidelity \cite{multiplexing_mimo}. For DV systems, multiplexing multiple heralded single photon sources can make a probabilistic single photon source behave more deterministically \cite{herald_multiplex_2026, herald_multiplex_2020, herald_multiplex_2013}. They can also make use of high-dimensional entanglement for denser information processing \cite{kai_chi_2025}. CV systems make use of large-scale entanglement to create cluster states, which is the core of measurement-based computations \cite{cv_computing_architecture_2025, large_scale_cluster_2025, microcomb_cluster_2025, cluster_2013}. Quantum imaging schemes like ghost imaging can use multiple photon pairs for alternative imaging schemes such as four-photon ghost imaging \cite{entangled_imaging_2019}.

Photon pairs can be created through multiple methods. Quantum dots can deterministically produce photon pairs via controlled excitations \cite{quantum_dot_2024}. A neutral atom can also produce photon pairs by coupling two cavities to two cascaded transitions \cite{neutral_atom_2024}. Nonlinear optical processes such as spontaneous four-wave mixing (SFWM) and spontaneous parametric down-conversion (SPDC) provide probabilistic photon pairs that can be generated at room temperature. SFWM and SPDC have the additional benefit of direct compatibility with nanophotonic platforms, where tightly-confining waveguides offer low propagation loss and nonlinear interactions at low powers compared to bulk optical platforms \cite{nanophotonics_for_quant_2024}. On nanophotonic platforms, SFWM and SPDC can be implemented with a quasi-phase-matched periodically poled region in either a waveguide or a resonant cavity. Waveguide implementations can achieve a continuous spectrum of photon pairs \cite{ln_wg_Fang_2024, ln_wg_Shi_2024, sn_wg_Choi_2020, shi2025_arxiv, fang2025_arxiv, kiwon_lol_2024, Zhao_2020}. In contrast, resonant cavities intrinsically perform spectral filtering and additionally enhance the nonlinear interaction time, alleviating design constraints on external filters and lowering input power requirements.

Photon pair sources based on resonant cavities have been demonstrated on various nanophotonic platforms, including SiN, InP, InGaP, AlGaAs, AlN, and LiNbO$_3$ \cite{SiN_2025, SiN_2022, SiN_2023, InP_ref, InGaP_ref, AlGaAs_ref, AlN_ref, z_cut_ln_ref}. Among various nanophotonic platforms, thin-film lithium niobate (LN) offers a strong intrinsic second-order optical nonlinearity ($\chi^{(2)}$), low propagation loss, and dispersion engineering for broadband frequency conversion \cite{ln_is_peak_Zhu_2021}. X-cut thin-film LN (TFLN) in particular, provides access to a mature platform for high-performing and low-loss EO modulators \cite{ln_modulators_review_2025, ln_modulators_nonclassical_2022, ln_modulators_2020}. These traits allow X-cut TFLN to be a platform for not only producing photon pairs, but also for manipulating photon-pairs for various applications.

In this work we demonstrate a multi-channel photon pair source from an X-cut TFLN microracetrack resonator with 80 measured channel pairs. Photon pairs are generated through SPDC in a periodically-poled segment in the resonator, which we design to be overcoupled at telecom resonances to enable more usable pairs in the bus waveguide. Pairs are produced in the optical C and L bands with a frequency spacing of 49.5 GHz, allowing them to be used with dense wavelength division multiplexing (DWDM). We measure the highest number of photon pair channels to date, enabled by the low FSR of our microresonator and the broad phase-matching bandwidth of our SPDC periodically-poled region. We also measure a high coincidence-to-accidental ratio of 37,004 $\pm$ 2,496 with an on-chip pump power of 357 nW, and confirm heralded single photon operation with a heralded correlation dip of 0.0544 $\pm$ 0.0054. Our work offers a platform to implement robust quantum applications using a high number of channel pairs and access to a mature nanophotonics ecosystem.

\section{Device Design and Fabrication}

The design procedure consists of selecting a material stack, determining parameters for the periodically-poled region, determining parameters for the resonant cavity, and engineering the bus-cavity coupling. A schematic of the device design and operating principle is depicted in Fig. 1a. The wafer stack is 5\% MgO-doped 600 nm X-cut LN on 2 μm SiO\textsubscript{2} on an Si support substrate. The MgO doping makes the LN more resistant to optical damage \cite{mgo_doping_2006}. The resonant cavity is a microracetrack resonator where one straight segment of the racetrack is periodically poled. The total ring length is 2668 μm, and the poled region is 0.7 mm with a poling period of 4.767 μm. The gap between the bus waveguide and microresonator is 0.7 μm, and the total coupling length is 15 μm. These parameters are chosen such that the external quality factor ($Q_e$) is less than the intrinsic quality factor ($Q_i$) at telecom wavelengths, resulting in an overcoupled coupling condition. The overcoupled condition ensures more photon pairs can escape into the bus waveguide rather than being lost in the microresonator.

The device fabrication follows a procedure of periodic poling, waveguide and microresonator etching, facet creation, and post-processing steps. First, round-tip comb electrodes consisting of 150 nm Au on top of 10 nm Cr were made by direct-write lithography followed by electron-beam evaporation. A thick layer of photoresist is used for insulation, as high-voltage pulses with millisecond durations were applied on the electrodes to pole the LN. After poling, the metal electrodes were removed via wet etching. Next, electron-beam lithography patterned the waveguide and microresonator with hydrogen-silsequioxane (HSQ) resist, which was followed by reactive-ion etching with Ar gas to etch 350 nm of LN. Afterwards, the wafer was cladded with SiO\textsubscript{2} through plasma-enhanced chemical vapor deposition (PECVD). To create chips, the wafer was diced, and the resulting chips were cleaved to form optical facets. An additional PECVD step was used to create anti-reflection coatings. Finally, the chip was annealed at 500 $^\circ$C for 5 hrs to improve the quality. A microscope image of the fabricated device and a second harmonic generation (SHG) confocal microscope image of the poling region is shown in Fig. 1b.

The device is characterized by its SHG efficiency, quality factors ($Q$s), and its downconverted spectrum. The SHG efficiency is obtained by pumping the chip with telecom light at 1560 nm, measuring the output visible power, and taking the ratio of output visible to the square input telecom power (Fig. 1c). The intrinsic and external quality factors are extracted to be $Q_i=1.3\times10^6$ and $Q_e=1.4\times10^7$ at visible wavelengths and $Q_i=1.8\times10^6$ and $Q_e=3.3\times10^5$ at telecom wavelengths (Fig. 1d and 1e). This results in an escape efficiency ($\eta_{esc}$) of 85\%. The downconverted spectrum is obtained by sending visible light into the device and measuring the output spectrum on an optical spectrum analyzer (OSA). To align telecom and visible resonances, the chip is mounted on a thermoelectric cooler (TEC), which modifies the refractive index through the thermo-optic effect. The chip is pumped with 26.2 mW of visible light near 780 nm, and the resulting downconverted spectrum spanning the optical C and L bands is shown in Fig. 1f. The dips and nonuniformity in the spectrum are mainly attributed to mode crossings between the fundamental and higher-order mode in the microresonator.

\section{Photon Pair Performance}

The performance of the device as a photon-pair source can be determined through cross-correlation and auto-correlation measurements of signal and idler photons. Correlation measurements reveal how likely one photon will be detected at a time relative to the detection of the other photon. By normalizing correlation measurements to obtain second-order coherence functions ($g^{(2)}(\tau)$), the photon statistics of the source are revealed.

\subsection{Cross-correlation}

\begin{figure}
\includesvg[width=0.5\textwidth]{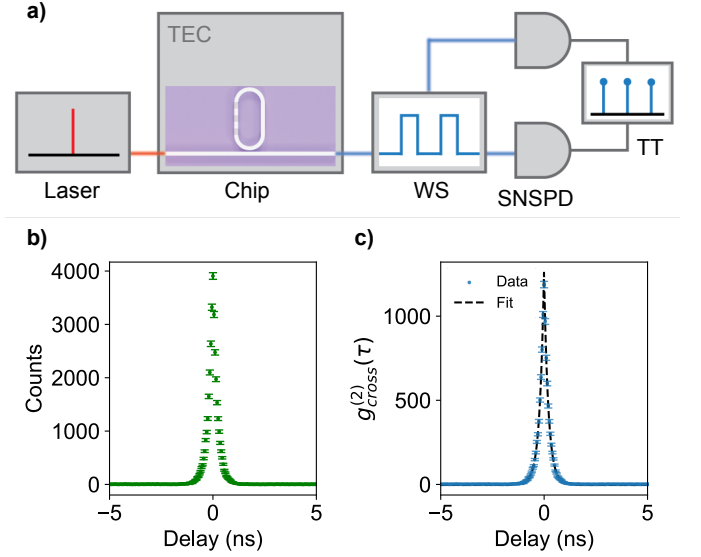}
\caption{\label{fig:epsart} Cross-correlation measurements. (a) Simplified measurement setup schematic. TEC: thermoelectric cooler; WS: waveshaper; SNSPD: superconducting nanowire single-photon detector; TT: time tagger (b) Raw time tagger counts with a 50 ps bin width and a 300 s measurement time of the 9th channel pair. (c) Normalized $g^{(2)}_{cross}(\tau)$ correlation function of the 9th channel pair.}
\end{figure}

\begin{figure*}
\includesvg[width=1\textwidth]{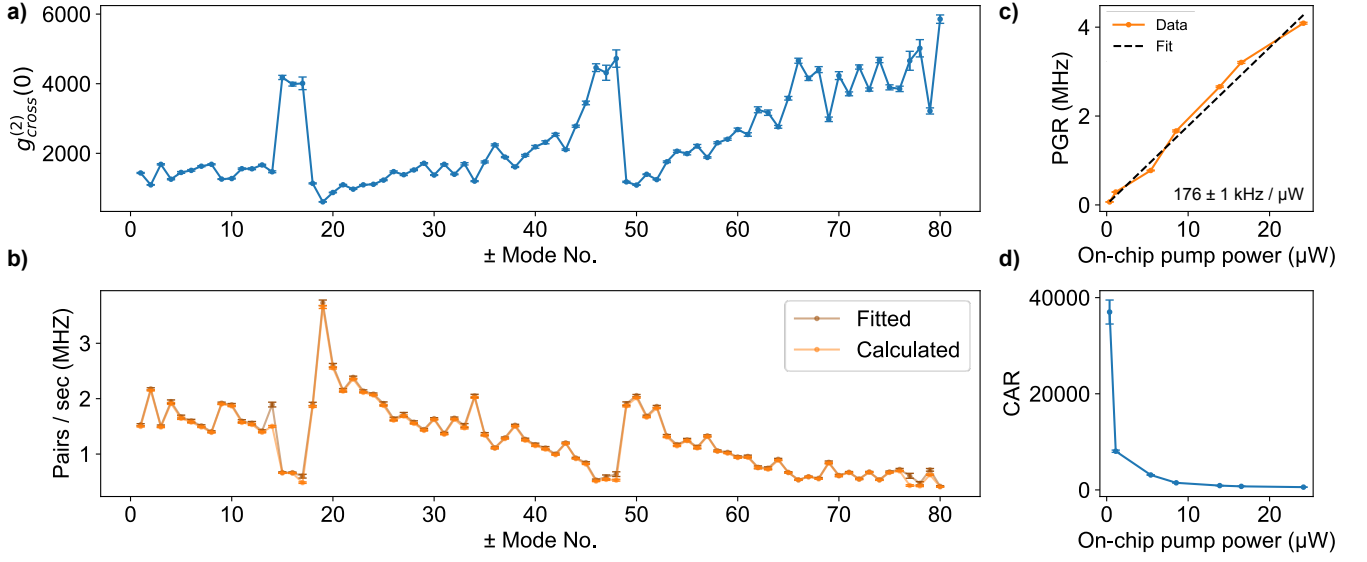}
\caption{\label{fig:full_characterization}Multi-mode and multi-power characterizations of the photon pair source. (a) $g^{(2)}_{cross}(0)$ values and (b) PGRs of 80 photon pair channels each measured for 300 s with 21 \mu W of on-chip pump power near 780 nm. The fitted and calculated PGRs show good agreement with each other. Calculated PGRs use a coincidence window of 2.2 ns. (c) PGRs (calculated) and (d) CAR values of the 9th channel pair at different pump powers.}
\end{figure*}

\begin{figure}
\includesvg[width=0.5\textwidth]{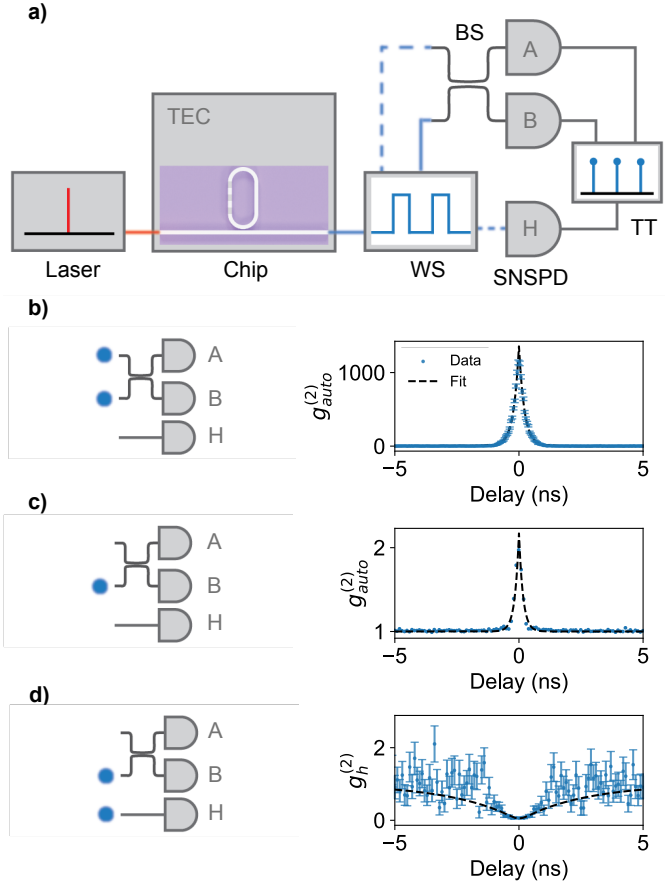}
\caption{\label{fig:auto_correlation}Auto-correlation measurements. (a) Simplified measurement setup schematic. TEC: thermoelectric cooler; WS: waveshaper; SNSPD: superconducting nanowire single-photon detector; TT: time tagger; BS: 50:50 beamsplitter (b) Frequency-degenerate schematic and $g^{(2)}_{auto}(\tau)$ histogram, using the same measurement settings as the non-degenerate correlation measurements. (c) Schematic and $g^{(2)}_{auto}(\tau)$ of the lower frequency photon from the 9th channel pair which exhibits thermal source statistics. The measurement is conducted using the same raw time tags as the heralded single photon measurement. (d) Heralded single photon schematic and $g^{(2)}_{h}(\tau)$ from 9th channel pair using a bin width of 100 ps, a coincidence window of 300 ps, a measurement time of 4 hours, and a visible pump power of 2.2 mW.}
\end{figure}

Cross-correlation measurements are performed using an experimental setup shown in Fig. 2a. Visible light near 780 nm pumps the TFLN chip mounted onto a TEC to generate the downconverted photon pair spectrum. A commercial waveshaper (Finisar WaveShaper 4000A) selects one photon pair channel by filtering out two cavity resonances corresponding to a signal-idler pair, and splits the signal and idler into two different physical paths based on their frequency. These paths are then sent to superconducting nanowire single photon detectors (SNSPDs) after filtering to attenuate residual visible pump and ambient light. The SNSPDs are connected to a time tagger for photon counting and correlation measurements. The measurement bin width is set to 50 ps to balance fine resolution and fast measurement time.

Eighty photon pair channels are measured in a continuous automated measurement lasting under 7 hours. The amount of channel pairs measured is limited by the waveshaper's frequency range. The chip input features an automated lensed fiber alignment system to mitigate input light coupling drift, while the chip temperature is stabilized with a PID loop via a thermoelectric cooler (TEC) controller. Each channel is measured for 300 s, and the chip uses 21 μW of on-chip pump power throughout the measurement. An example of a raw count correlation measurement is shown in Fig. 2b using the 9th channel pair, which corresponds to the 9th cavity resonance before and after the degenerate cavity resonance. The degenerate cavity resonance refers to the resonance that aligns with frequency-degenerate photon pairs at half the visible pump light frequency. Error bars are determined by Poissonian counting statistics. From the raw counts, the normalized $g^{(2)}_{cross}(\tau)$ correlation function can be obtained and is given by

\begin{equation}
    g^{(2)}_{cross}(\tau)=\frac{T}{N_AN_B\Delta t}\times\text{histogram}(\tau),
\end{equation}

where $T$ is the coincidence measurement time, $\Delta t$ is the bin width, and $N_A$ and $N_B$ are total counts on each detector. The correlation is normalized to a coherent source where the counts are evenly distributed across all time bins and have a normalized $g^{(2)}_{cross}(\tau)$ of 1. The $g^{(2)}_{cross}(\tau)$ function can then be fitted by 

\begin{equation}
    g^{(2)}_{cross}(\tau)=1+\frac{1}{2R\tau_c}e^{-|\tau|/\tau_c},
\end{equation}

where $R$ is the pair generation rate (PGR) inside the nonlinear interaction region, and $\tau_c$ is the coherence time of the photons \cite{AlN_ref}. By fitting the $g^{2}_{cross}(\tau)$ function, $g^{(2)}_{cross}(0)$ is extracted to be 1260.2 $\pm$ 9.8 and the coherence time of the photons is extracted to be 206.9 $\pm$ 1.2 ps (Fig. 2c), where the uncertainties are one standard deviation from the mean.

The characterization of 80 measured photon pair channels is shown in Fig. 3a. The trend in the $g^{(2)}_{cross}(0)$ values aligns with the downconverted spectrum (Fig. 1f), where peaks in $g^{(2)}_{cross}(0)$ values corresponds to fundamental and higher-order mode crossings in the microresonator. The $g^{(2)}_{cross}(0)$ values are also directly related to the coincidence-to-accidental ratio (CAR) by $\text{CAR}=g^{(2)}_{cross}(0)-1$ \cite{photon_measurements_2020}. While fitting $g^{(2)}_{cross}(\tau)$ can extract PGR, PGR can also be calculated using 

\begin{equation}
    \text{PGR}=\frac{R_{s}R_{i}}{R_{si}},
\end{equation}

where $R_s$ is the detected signal photon count rate, $R_i$ is the detected idler photon count rate, and $R_{si}$ is the coincidence rate of the signal and idler. Considering the photon coherence time, a coincidence window of 2.2 ns is chosen for measuring coincidence count rates to capture most coincidence events. The fitted and calculated PGRs are presented in Fig. 3b. PGR and CAR are also measured at different on-chip pump powers for the 9th channel pair, and are shown in Fig. 3c and Fig. 3d, respectively. The PGR varies at the same power in the 80 channel pair sweep and the power sweep due to different coupling conditions at the chip input. Through a linear fit, the PGR vs pump power is determined to be 176 $\pm$ 1 kHz / μW. There is a fundamental tradeoff between PGR and CAR due to the higher probability of multi-pair emission at higher pump powers, which increases the accidental coincidence rate.

\subsection{Auto-correlation}

\begin{figure}
\includesvg[width=0.5\textwidth]{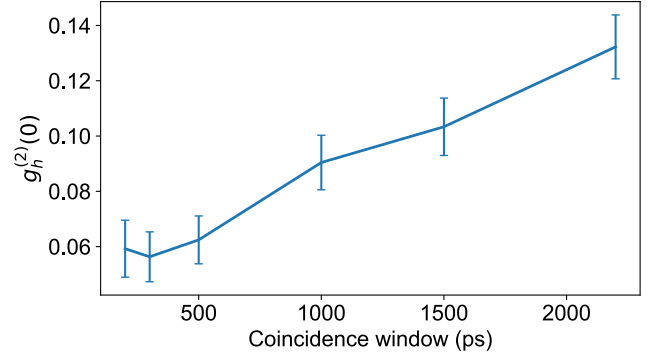}
\caption{\label{fig:auto_correlation}Coincidence window's effect on calculated heralded $g^{(2)}_{h}(0)$. All data points are obtained using the same raw time tags, leading to all error bars being the same.}
\end{figure}

Auto-correlation measurements can characterize the frequency-degenerate photon pair channel, verify thermal state statistics, and observe heralded single photons. The experimental setup is modified by adding a 50:50 beamsplitter to either the signal or idler path after the waveshaper and incorporating an additional detector. Performing a correlation measurement on the two detectors after the beamsplitter (detectors $A$ and $B$) measures the auto-correlation of the light entering the beamsplitter. The experimental schematic for auto-correlation is shown in Fig. 4a. The degenerate photon pairs are measured by filtering only the degenerate resonance and routing the photons to the beamsplitter. When fitting the $g^{(2)}_{auto}(\tau)$ of degenerate pairs, there is an additional factor of $1/2$ in the second term since there is a 50\% chance that both photons are incident on the same detector. The function for fitting the degenerate pairs is given by

\begin{equation}
    g^{(2)}_{auto}(\tau)=1+\frac{1}{4R\tau_c}e^{-|\tau|/\tau_c}.
\end{equation}

The extracted $g^{(2)}_{auto}(0)$ and PGR are 1360.2 $\pm$ 16.7 and 848.7 $\pm$ 12.8 kHz, respectively, and the correlation histogram is presented in Fig. 4b. The frequency-degenerate photon pair auto-correlation is measured with the same measurement settings as the non-degenerate photon pair cross-correlations. The statistics of one photon in a photon pair generated from SPDC are expected to be thermal, which has a $g^{(2)}(0)$ value of 2. To characterize this, the waveshaper selects the 9th channel pair, and the auto-correlation of the lower frequency photon is measured. Fig. 4c shows the resulting $g^{(2)}_{auto}(\tau)$, with a $g^{(2)}_{auto}(0)$ value of 2.173 $\pm$ 0.015, which agrees with the $g^{(2)}(0)$ value of a thermal state. By using the photon incident on detector $H$ as a herald, anti-bunching statistics can be observed where $g^{(2)}(0)<1$, indicating a single photon source. The heralded $g^{(2)}_{h}(\tau)$ is given by

\begin{equation}
    g^{(2)}_{h}(\tau)=\frac{R_{ABH}(\tau)R_H}{R_{AH}R_{BH}(\tau)},
\end{equation}

where $R_{ABH}(\tau)$ is a correlation between photons incident on the $B$ detector and coincidences between heralding photons and photons incident on the $A$ detector, $R_H$ is the count rate of heralding photons, $R_{AH}$ is the coincidence count rate of photons incident on the $A$ detector and heralding photons, and $R_{BH}(\tau)$ is the coincidence count rates between photons incident on the $B$ detector and heralding photons. The calculated heralded $g^{(2)}_{h}(0)$ value is 0.0544 $\pm$ 0.0054 using a coincidence window of 300 ps (Fig. 4d). The coincidence window is chosen where its effect on $g^{(2)}_{h}(0)$ begins to saturate \cite{coherence_measures_2009}. The dependence of the heralded $g^{(2)}_{h}(0)$ value on the coincidence window width can be obtained by re-running the measurement on the same raw time tags, but changing the coincidence window width. This effect is shown in Fig. 5 while maintaining a bin width of 100 ps.

The measurements for the heralded single photon and the thermal state verification can be performed on the same raw time tags, thus have the same measurement settings. Since only two C-band SNSPDs were available in the experiment, an O-band SNSPD was used as the heralding detector with an efficiency of 1\% to 3\% across 1520 nm to 1600 nm. Due to a low heralding photon rate, the bin width was increased to 100 ps, and the measurement time was increased to 4 hours to collect a satisfactory number of counts. With a long measurement time, the error for the thermal state verification becomes negligible, and the error bars can be omitted. Additionally, the visible pump power is increased to 2.2 mW to increase the PGR for combating the low detector efficiency. Nevertheless, a $g^{(2)}(0)$ value below 1 is a characteristic of photon anti-bunching, and the closer $g^{(2)}(0)$ is to zero, the closer the source is to an ideal single photon source.

\section{Discussion and Conclusion}

\begin{table*}
\caption{\label{tab:table3}Microresonator photon pair platforms comparison}
\begin{ruledtabular}
\begin{tabular}{ccccccccc}
 Platform & Process & Telecom $Q_i$ & $\eta_{esc}$ & \makecell{PGR $\cdot$ $\eta_{esc}^2$ / μW (SPDC) \\ PGR $\cdot$ $\eta_{esc}^2$ / μW\textsuperscript{2} (SFWM)} & \makecell{Highest \\ CAR} & \makecell{Highest CAR \\ pump power} & \makecell{Measured number \\ of channel pairs} & \makecell{FSR \\ (GHz)}\\ \hline
SiN\cite{SiN_2022} & SFWM & n/a & n/a & n/a & 137 & 640 μW & 6 & 97.8 \\
SiN\cite{SiN_2023} & SFWM & $3.5\times10^6$ & 71\% & 76.5 mHz / μW$^2$ & 1,243 & 250 μW & 7 & 200 \\
InP\cite{InP_ref} & SFWM & n/a & n/a & n/a & 277 & 22 μW & 1 & 421 \\
InGaP\cite{InGaP_ref} & SPDC & $1.75\times10^5$ & 39\% & 4.1 MHz / μW & 1400 & 5.5 nW & \makecell{Degenerate \\ only} & n/a \\
AlGaAs\cite{AlGaAs_ref} & SFWM & n/a & n/a & $\leq$ 20 kHz / μW$^2$ & 4,389 & 3.39 μW & 1 & 927 \\
AlN\cite{AlN_ref} & SPDC & $2.7\times10^6$ & 93\% &  5.0 kHz / μW & 560 & 600 μW & 1 & 749 \\
Z-cut LN\cite{z_cut_ln_ref} & SPDC & $2.6\times10^5$ & 62\% & 1.0 MHz / μW & 24,583 & 210 nW & 6 & 366 \\
\textbf{X-cut LN} & \textbf{SPDC} & \textbf{1.8 × 10\textsuperscript{6}} & \textbf{85\%} & \textbf{125.7 kHz / μW}\footnotemark[1] & \textbf{37,004}\footnotemark[1] & \textbf{357 nW} & \textbf{80} & \textbf{49.5} \\
\end{tabular}
\end{ruledtabular}
\footnotetext[1]{Measured with channel 9}
\end{table*} 

Previous demonstrations of photon pair sources based on microresonators using SPDC have been able to achieve high pair generation rates; however, they are limited in their multi-channel use due to relatively large FSRs, especially considering DWDM standards. A comparison of microresonator-based photon pair sources is provided in Table I. When comparing the pair production of microresonator-based photon pair sources, the PGR per pump power is multiplied by $\eta_{esc}^2$ to get the on-chip generation rate, as two photons need to exit the resonator to become a usable photon pair. We report the measured channel 9 on-chip generation rate as 125.7 kHz / μW, and use the relative PGRs in Fig. 3b to obtain the highest on-chip generation rate of 238.8 kHz / μW with channel 19 and an average on-chip generation rate of 83.5 kHz / μW over all 80 channel pairs. SPDC sources linearly depend on pump power, and SFWM sources depend quadratically on the pump power because a three-wave-mixing conversion requires one pump photon and a four-wave-mixing conversion requires two pump photons. The highest CAR values are reported alongside the on-chip pump power used, since the on-chip pump power can be reduced to improve the CAR. In addition to channels characterized, available channel pairs are inferred from transmission. Although not the brightest source in terms of PGR $\cdot$ $\eta_{esc}^2$ / μW, our X-cut LN source is still competitive amongst other platforms and possesses a far greater amount of channel pairs compared to previously demonstrated sources.

In conclusion, we have experimentally measured 80 frequency-correlated photon pairs from a periodically-poled racetrack microresonator on X-cut TFLN. We created an overcoupled resonator at telecom wavelengths by engineering the coupling between the bus waveguide and resonator to extract more usable photon pairs from the resonator. We characterized the photon pair source by $g^{(2)}(\tau)$ correlation functions, PGR, and CAR for non-degenerate photon pairs. We also measured the frequency of degenerate photon pairs, confirmed thermal state statistics of a single photon from a pair, and verified the ability of the source to function as a heralded single photon source via a heralded single photon autocorrelation. These results demonstrate X-cut TFLN as a promising platform for scalable quantum applications with the benefit of including competitive quantum sources and fast low-loss modulators all on a single chip, which can have applications in quantum information processing, communication, and quantum sensing.

\begin{acknowledgments}
This work is supported by the DARPA INSPIRED program (HR001123S0052) and the DARPA Young Faculty Award (D23AP00252-02). We thank Single Quantum for providing SNSPDs and Swabian Instruments for providing a time tagger. Two-photon imaging was conducted at the CRL Molecular Imaging Center, RRID: SCR017852, supported by NIH grant S10OD025063. Device fabrication was performed at the John O’Brien Nanofabrication Laboratory at University of Southern California and Marvell Nanofabrication Laboratory at University of California, Berkeley. M.Y. and T.K. are supported by the U.S. Department of Energy, Office of Science, Basic Energy Sciences, Materials Sciences and Engineering Division under Contract No. DE-AC02-05CH11231 within the Quantum Coherent Systems Program KCAS26. The views, opinions and/or findings expressed are those of the authors and should not be interpreted as representing the official views or policies of the Department of Defense or the U.S. Government.
\end{acknowledgments}

\section*{Author Declarations}

\subsection*{Conflict of Interest}
C.-H.L., and M.Y. are involved in developing lithium niobate technologies at Opticore Inc.

\subsection*{Author Contributions}
\textbf{M.C.} Measurement (lead); Data analysis (lead); Writing - original draft (lead). \textbf{I.C.} Measurement (supporting); Data analysis (supporting). \textbf{X.R.} Measurement (supporting). \textbf{C.L.} Device design (lead); Device fabrication (lead). \textbf{T.K.} Device poling (equal). \textbf{R.K.} Device poling (equal). \textbf{C.C.} Device fabrication (supporting). \textbf{K.C.} Writing - original draft (support) \textbf{M.Y.} Conceptualization (lead); Supervision (lead); Writing - review \& editing (lead).

\section*{Data Availability}

The data that support the findings of this study are available from the corresponding author upon reasonable request.

\section*{References}
\nocite{*}
\bibliography{aipsamp}% Produces the bibliography via BibTeX.

\end{document}